\documentclass[aps,twocolumn,groupedaddress,showpacs,floats,amssymb]{revtex4}
\usepackage[dvips]{graphicx}
\usepackage{bm}

\begin{document}

\noindent {\bf Comment on ``Hausdorff Dimension of Critical
Fluctuations in Abelian Gauge Theories"}\vspace{2mm}\\

In their Letter \cite{Hove},  Hove, Mo, and Sudb{\o} derive a
simple connection between the anomalous scaling dimension, $\eta$,
of the U(1) universality class order parameter, $ \phi({\bf x})$,
and the Hausdorff dimension, $D_H$, of critical loops:
\begin{equation}
\eta + D_H = 2 \;  . \label{wrong}
\end{equation}
In the loop representation, the correlator $G({\bf r}) = \langle
\phi({\bf r}) \phi^*(0) \rangle\, \propto\, r^{-(d-2 +\eta )}$
describes the distribution of the end-points in open loops. For
definiteness, one may think of the high-temperature-expansion
loops for the lattice $|\phi|^4$-model.

The analysis of Ref.~\cite{Hove} might seem absolutely compelling,
being just a translation of the hyperscaling hypothesis into the
loop language: {\it At the critical point there should be about
one loop of diameter $r$ per volume element $r^d$}
\cite{Williams}. Nevertheless,  given the result $\eta =
0.0380(4)$ of Ref.~\cite{Campostrini}, the relation (\ref{wrong})
is in strong contradiction with the value $D_H=1.7655(20)$ which
we obtained for the 3D $|\phi|^4$-model with suppressed leading
corrections to scaling \cite{Campostrini} (and also---with a bit
less accuracy---for the standard bond-current model \cite{Wallin},
and its special version with excluded loop overlaps and
self-crossings). The simulations were done with the Worm algorithm
\cite{Worm}.

The hidden flaw in the treatment of Ref.~\cite{Hove} is as
follows. When introducing the self-similar expression
\begin{equation}
P({\bf r};N)\; \propto \; N^{-\rho}\, F(r/N^\Delta )\; , ~~~
\Delta =  1/D_H  \; \label{P}
\end{equation}
for the probability to find the ends of an open loop of length $N$
being distance  ${\bf r}$ away from each other, which is then used
to establish the connection between the open and closed loops, the
authors take for granted that $F(0)$ is finite. While looking
innocent, this is an {\it arbitrary} assumption, since the
self-similar form (\ref{P}) is valid only for $r \gg a$, where $a$
is a microscopic cutoff (e.g., the lattice period). Strictly
speaking, a closed loop of length $N$ corresponds to $F(a/N^\Delta
)$ rather than to $F(0)$, and one has to work with the generic
asymptotic form
\begin{equation}
F(x) \, \propto\,  x^{\theta}~~~~~\mbox{at}~~~~~x \ll 1 \; ,
\label{theta}
\end{equation}
with some exponent $\theta$. With Eq.~(\ref{theta}), the
hyperscaling argument yields $\rho=(d-\theta)/D_H$, and from
$G(r)\propto \int dN P({\bf r};N)$ one then obtains
\begin{equation}
\eta + D_H = 2-\theta \; . \label{correct}
\end{equation}
Using high-precision data for $\eta$ and $D_H$ mentioned above, we
find $\theta = 0.1965(20)$.

\begin{figure}[htb]
\includegraphics[bb=0 100 600 700, angle=-90, width=0.68\columnwidth]{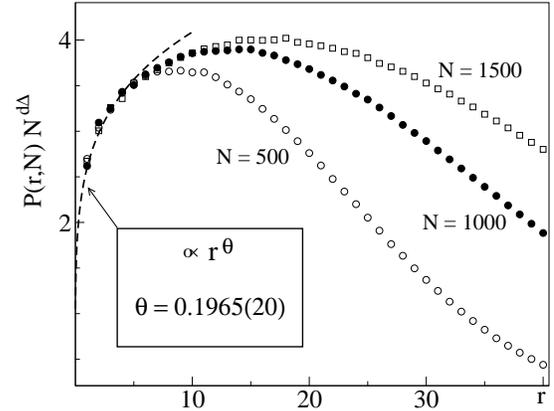}
\caption{Distribution of open loops over radii for three different
values of $N$. The Worm algorithm simulation \cite{Worm} was done
for the loop representation (high-temperature expansion) of the 3D
lattice $|\phi|^4$-model with $L=192^3$ sites at the special
critical point with suppressed leading corrections to scaling
\cite{Campostrini}.
 } \label{Fig1}
\end{figure}

It is instructive to explicitly verify Eq.~(\ref{theta}) by
simulating $P(r;N)$. In Fig.~\ref{Fig1} we present results of such
a simulation for the $|\phi|^4$-model. We plot the value of
$P(r,N)N^{d\Delta}$ as a function of $r$ for three different
values of $N$. In view of the self-similarity of $P(r,N)$, the
qualitative difference between the cases of $\theta \neq 0$ and
$\theta = 0$ is readily seen. In the former case, curves for
different values of $N$ should merge for $r/N^{\Delta} \ll
1$---and they do in Fig.~\ref{Fig1}. In the latter case, as $r \to
0$ one should see a fan of curves with essentially different
slopes and a common origin at $r=0$.

One important implication of Eq.~(\ref{correct}) in the absence of
additional relation between $D_H$, $\eta $ and $\theta$,  is that
the anomalous scaling dimension {\it can not} be deduced from
simulations of closed loops which determine $D_H$ only.

We are grateful to A.~Kuklov, Z. Te\v{s}anovi\'{c}, and G.
Williams for numerous discussions. This work was supported by the
National Science Foundation under Grant
DMR-0071767.\vspace{2mm}\\
{\small \noindent  Nikolay Prokof'ev and Boris Svistunov,\\
Department of Physics,  University of Massachusetts, Amherst, MA
01003.}
\bigskip


\end{document}